\documentclass[aps,prx,twocolumn,superscriptaddress,longbibliography]{revtex4-1}

\usepackage{amsmath}
\usepackage{amssymb}
\usepackage{graphicx}
\usepackage[dvipsnames]{xcolor}
\usepackage[colorlinks,linkcolor=blue,citecolor=blue,urlcolor=blue]{hyperref}
\usepackage{stix}

\begin{document}

\renewcommand{\vec}[1]{\mathbf{#1}}
\newcommand{\iu}{\mathrm{i}}
\newcommand{\hc}{\hat{c}}
\newcommand{\hcd}{\hat{c}^\dagger}
\newcommand{\en}{\varepsilon}
\newcommand{\gvec}[1]{\boldsymbol{#1}}
\renewcommand{\pl}{\parallel}

\newcommand{\new}[1]{\textcolor{WildStrawberry}{#1}}

\author{Michael Sch\"uler}
\email{michael.schueler@psi.ch}
\affiliation{Condensed Matter Theory Group, Paul Scherrer Institute, CH-5232 Villigen PSI, Switzerland}
\author{Samuel Beaulieu}
\affiliation{Université de Bordeaux - CNRS - CEA, CELIA, UMR5107, F33405, Talence, France}

\title{Probing Topological Floquet States in WSe$_2$ using Circular Dichroism in Time- and Angle-Resolved Photoemission Spectroscopy}


\begin{abstract}
Observing signatures of light-induced Floquet topological states in materials has been shown to be very challenging. Angle-resolved photoemission spectroscopy (ARPES) is well suited for the investigation of Floquet physics, as it allows to directly probe the dressed electronic states of driven solids. Depending on the system, scattering and decoherence can play an important role, hampering the emergence of Floquet states. Another challenge is to disentangle Floquet side bands from laser-assisted photoemission (LAPE), since both lead to similar signatures in ARPES spectra. Here, we investigate the emergence of Floquet state in the transition metal dichalcogenide $2H$-WSe$_2$, one of the most promising systems for observing Floquet physics. We discuss how the Floquet topological state manifests in characteristic features in the circular dichroism in photoelectron angular distributions (CDAD) that is determined by the transient band structure modifications and the associated texture of the orbital angular momentum. Combining highly accurate modeling of the photoemission matrix elements with an \emph{ab initio} description of the light-matter interaction, we investigate regimes which can be realized in current state-of-the-art experimental setups. The predicted features are robust against scattering effects and are expected to be observed in forthcoming experiments.
\end{abstract}

\maketitle


\section*{Introduction}

Controlling properties of quantum materials by tailored light is at the forefront of condensed matter physics~\cite{basov_towards_2017,de_la_torre_nonthermal_2021} due to recent advances of ultrafast laser technologies.
In particular, manipulating the electronic structure's topology and creating on-demand topological properties -- theoretically predicted in ref.~\cite{oka_photovoltaic_2009} -- is one of the most fundamental goals of the field. The key idea is that periodically driven solids (described by Floquet theory) form effective bands that correspond to a topologically non-trivial state. While countless theoretical proposals exist, direct experimental observations of Floquet physics are scarce. The first experiments confirming the Floquet hallmarks -- emergence of side bands and gap openings -- have been performed on the surface of Bi$_2$Se$_3$~\cite{wang_observation_2013,mahmood_selective_2016} using time- and angle-resolved photoemission spectroscopy (trARPES). Observing Floquet features in graphene, following the original proposal from ref.~\cite{oka_photovoltaic_2009}, has proven difficult, as the interplay of decoherence~\cite{sato_floquet_2019,sato_microscopic_2019-1} and scattering effects~\cite{schuler_how_2020,gierz_tracking_2015,aeschlimann_survival_2021-1} is adversarial to the formation of Floquet states. Indeed, it has been concluded that Floquet-Bloch states cannot emerge if the scattering time of the electrons is shorter than or comparable to the period of the driving field~\cite{aeschlimann_survival_2021-1}. Furthermore, laser-assisted photoemission (LAPE)~\cite{miaja-avila_laser-assisted_2006} often overshadows the Floquet side bands~\cite{mahmood_selective_2016,keunecke_electromagnetic_2020}. To enhance the typically weak photodressing of the electronic bands, using intense low-frequency pump pulses emerged as a new direction~\cite{aeschlimann_survival_2021-1,broers_detecting_2021} to investigate Floquet physics in real systems. However, in this regime, scattering effects are becoming more pronounced: if the duration of the optical cycles is long compared to the decoherence time, the emergence of Floquet features is suppressed. Alternatively, evidencing the Floquet-topological state in light-driven graphene via transport measurements~\cite{mciver_light-induced_2019,broers_observing_2021-1} is a promising route, albeit the manifestation of the topology in ref.~\cite{mciver_light-induced_2019} is far from clear~\cite{sato_microscopic_2019-1}. 

\begin{figure}[b]
\centering\includegraphics[width=0.9\columnwidth]{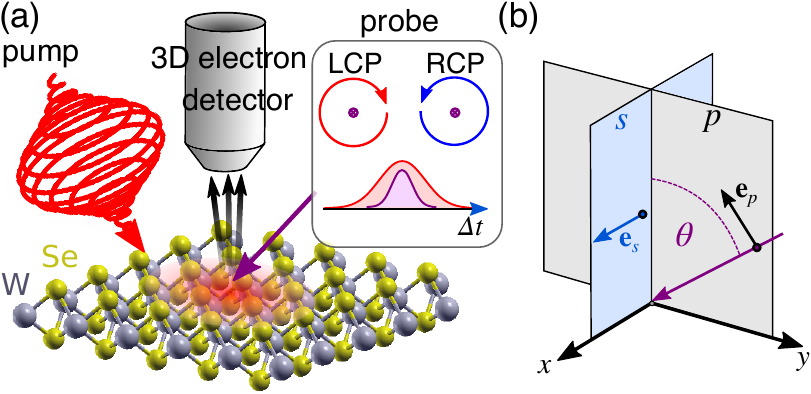}
\caption{\textbf{Sketch of the setup.} (a) A circularly polarized pump pulse transiently dresses the electronic structure of bulk $2H$-WSe$_2$, which is probed by a circularly polarized femtosecond XUV pulse. The probe pulse is centered at time delay $\Delta t=0$ with respect to the peak of the pump pulse. (b) Geometry of the simulated experiment: the propagation direction of the probe pulse is in the $p$ plane ($x$--$y$ plane), rotated by an angle of $\theta=65^\circ$ with respect to the $z$ axis. The $s$ plane is along the $x$ axis.}
\label{fig:sketch}
\end{figure}

trARPES is the most direct experimental technique to access the electronic structure and occupations in photodressed solids~\cite{smallwood_ultrafast_2016,rohde_time-resolved_2016,lee_high_2020,maklar_quantitative_2020}. However, instead of focusing on the spectral features of photodressed band structure alone, mapping out properties of the associated Bloch wave-functions would yield a new level of insight. Information on the Bloch wave-function manifests in the complex photoemission matrix elements (interference effects and anisotropy of the photoemission intensity~\cite{moser_experimentalists_2017,moser_huygens_2022}), leading to (linear and circular) dichroism in the ARPES spectrum. In static ARPES, the circular dichroism in the photoelectron angular distributions (CDAD) has been proven to be a powerful tool to map out pseudospin properties~\cite{liu_visualizing_2011,gierz_graphene_2012}, helical spin textures~\cite{wang_observation_2011,lin_orbital-dependent_2018,jozwiak_spin-polarized_2016-1}, high-symmetry planes~\cite{fedchenko_4d_2019} in 3D reciprocal-space, and Berry curvature~\cite{razzoli_selective_2017,cho_experimental_2018,cho_studying_2021,schuler_local_2020-1,unzelmann_momentum-space_2021}. The sensitivity of the CDAD to topological properties is due to the intimate connection between Berry curvature and orbital angular momentum (OAM)~\cite{xiao_berry_2010}. This link provides a new avenue for observing light-induced topology from features in the CDAD in trARPES.

While graphene is the paradigmatic material for the emergence of light-induced topological states~\cite{oka_photovoltaic_2009,schuler_how_2020}, complications like the sample size and/or quality as well as very fast electronic scattering time, render such experimental demonstration very difficult~\cite{aeschlimann_survival_2021-1}. At variance, Floquet physics has been predicted by first-principle calculations~\cite{de_giovannini_monitoring_2016} and has been observed experimentally in transition metal dichalcogenides (TMDCs)~\cite{liu_femtosecond_2017,aeschlimann_survival_2021-1}. Furthermore, as predicted in ref.~\cite{claassen_all-optical_2016-1}, a Floquet-Chern insulating state can be realized in the conduction band manifold. This raises the question: can we use CDAD in time-resolved ARPES to directly probe the light-induced topological state in driven TMDCs?

In this work, we address this fundamental question by predictive simulations. We are motivated by recent experimental progress that allows to perform XUV-trARPES measurements with femtosecond time resolution~\cite{maklar_quantitative_2020,sie_time-resolved_2019,corder_ultrafast_2018,buss_setup_2019,cucini_coherent_2020-1,mills_cavity-enhanced_2019,puppin_time-_2019,chiang_boosting_2015,keunecke_time-resolved_2020}. Achieving circularly polarized XUV pulses at hundreds of kilohertz repetition-rate is the key technical challenge for measuring transient CDAD across the entire Brillouin zone. While this was not demonstrated yet, all the technological building blocks (high-repetition-rate XUV source and XUV quarter--wave plate~\cite{dohring_circular_1992,vodungbo_polarization_2011,yao_tabletop_2020,von_korff_schmising_generating_2017}) allowing to performed such experiments are available and are currently being put together by some experimental groups (including one of the authors). The purpose of this paper is to predict the Floquet features in laser-driven TMDC and the associated CDAD, under conditions that are directly compatible with trARPES setups. 

Our main finding is that the emergence of Floquet side bands is accompanied by the formation of specific orbital textures with locally non-trivial topological characters, giving rise to a distinct CDAD signal. We also show that the dichroic signal is robust against dissipation and should be observable in a wide range of parameters. 
We use atomic units (a.u.) unless stated otherwise.

\section*{Results}

Analogous to the previous experiments~\cite{beaulieu_revealing_2020-1,schuler_polarization-modulated_2021} we consider the surface of bulk $2H$-WSe$_2$. To study the transient photodressing effects, we consider the following pump-probe setup. A strong, circularly polarized pump pulse (typical peak intensity in the range of $I_0 \sim 10^{11}$\, W/cm$^2$) induces Floquet features in the electronic structure, which are probed by circularly polarized XUV probe pulses (see Fig.~\ref{fig:sketch}(a)). Similar to typical ARPES setups using time-of-flight detector allowing to record the photoemission signal in the entire Brillouin zone within a single measurement~\cite{madeo_directly_2020,keunecke_time-resolved_2020,beaulieu_revealing_2020,beaulieu_ultrafast_2021,dong_direct_2021}, we assume an incidence angle of $\theta=65^\circ$ for the propagation direction of the XUV probe, as illustrated in Fig.~\ref{fig:sketch}(b). Varying the time delay $\Delta t$ between the pump and probe pulses allows observing the transient build-up of the photodressing effects. However, the Floquet features in the trARPES spectrum are maximized for the probe pulse centered at the peak of the envelope of the pump pulse, which is the regime we focus on in this work. In a typical setup, both the pump and the probe pulse impinge under the angle $\theta$. However, for simplicity we take the circular polarization of the pump to be in the $x$--$y$ plane, which can effectively be implemented by generated elliptically polarized pump pulses propagating collinearly with the probe pulse. The out-of-plane component does not play an important role for $2H$-WSe$_2$ due to the weak coupling of the layers. The results presented in this work are also robust against lifting this restriction.

\subsection*{Model and light-matter coupling}

As discussed in ref.~\cite{beaulieu_revealing_2020-1}, in the above described experimental condition, only the top-most layer of the bulk sample contributes notably to the photoemission signal. Hence, for our theoretical investigations, we model the system by a monolayer of WSe$_2$, which simplifies the theory considerably while retaining the relevant physics. The electronic properties and associated photoemission intensities in the vicinity of the K, K$^\prime$ valleys are well captured by a monolayer; small corrections to the photoemission properties due to bilayer interference -- as discussed in ref.~\cite{rostami_layer_2019} -- are built into the theory by fitting to experimental data, as explained below in the method section. 

The electronic structure is described by the first-principle Wannier model from ref.~\cite{schuler_gauge_2021-1}, which comprises the Se $p$ orbitals and W $d$ orbitals (11 orbitals in total). In Wannier presentation, the Bloch states of band $\alpha$ are given by 
\begin{align}
	\label{eq:wanrep}
	\psi_{\vec{k}\alpha}(\vec{r}) = \frac{1}{\sqrt{N}}\sum_{\vec{R},j}e^{i \vec{k}\cdot \vec{R}} C_{j\alpha}(\vec{k}) \phi_j(\vec{r}-\vec{R}) \equiv \sum_j C_{j\alpha}(\vec{k}) \varphi_{\vec{k}j}(\vec{r})  \ ,
\end{align}
where $\vec{R}$ runs over all $N$ lattice sites; $\phi_j(\vec{r})$ denote the Wannier orbitals. The coefficients $C_{j\alpha}(\vec{k})$ determine the complex superposition of the orbitals and thus the orbital texture. 
To describe the photoemission by the XUV probe pulse, we need the photoemission matrix elements $M_\alpha(\vec{k},E) = \langle \vec{k}, E| \vec{e}\cdot \hat{\vec{p}} | \psi_{\vec{k}\alpha} \rangle$ ($\hat{\vec{p}}$ denotes the momentum operator) with respect to the initial Bloch state and the final state determined by the in-plane momentum $\vec{k}$ and the energy of the photoelectron $E$.
Matrix element effects encode the orbital texture and the orbital angular momentum of the Bloch states. For our simulations to be predictive, an accurate model of matrix elements is thus crucial. We construct such a model by directly fitting calculated photoemission intensities to experimental data, which contains characteristic signal modulation due to matrix element effects. The most pronounced feature close to the valence band maximum (VBM) is the so-called dark corridor, i.\, e. the suppression of intensity along the edges of the Brillouin zone. The dark corridor is directly related to the interference of the relevant orbitals close to the VBM~\cite{rostami_layer_2019,beaulieu_revealing_2020-1} and is very sensitive to both the underlying orbital character $\phi_j(\vec{r})$ and their relative phase~\cite{schuler_polarization-modulated_2021}. Therefore, a model that contains the correct orbital symmetries and reproduces the orientation of dark corridors in the Brillouin zone contains all the relevant phase information ingredients to predict spectra for both linear and circular polarization~\cite{schuler_polarization-modulated_2021}. We construct such a model from the ansatz
\begin{align}
	\label{eq:mel_ansatz}
	M_\alpha(\vec{k},E) = \sum_j C_{j\alpha}(\vec{k})e^{i\gamma_j} M^{\mathrm{(at)}}_j(\vec{k},E) \ ,
\end{align}
where $M^{\mathrm{(at)}}_j(\vec{k},E)$ denote the matrix elements with respect to the Wannier orbitals, calculated similarly as in atomic physics, while $\gamma_j$ are additional phase shifts that account for final state effects. We have fitted the phases $\gamma_j$ and the few parameters entering $M^{\mathrm{(at)}}_j(\vec{k},E)$ to experimental spectra from ref.~\cite{schuler_polarization-modulated_2021} for both $s$ and $p$ polarized light (see methods section Supplementary Note for details and the explicit comparison).

The Wannier representation~\eqref{eq:wanrep} also provides a direct path to computing the light-matter coupling matrix elements that allow us to incorporate the pump field in a first-principle fashion~\cite{schuler_gauge_2021-1}. Here we adopt the velocity gauge in which the time-dependent Hamiltonian (including the pump pulse only) is given by
\begin{align}
	\label{eq:tdham}
	H_{\alpha\alpha^\prime}(\vec{k},t) = \left(\en_\alpha(\vec{k})+\frac{\vec{A}_\mathrm{p}(t)^2}{2}\right) \delta_{\alpha \alpha^\prime} - \vec{A}_\mathrm{p}(t)\cdot \vec{v}_{\alpha \alpha^\prime}(\vec{k}) \ ,
\end{align}
where $\en_\alpha(\vec{k})$ is the band energy of corresponding the Bloch state, while $\vec{v}_{\alpha \alpha^\prime}(\vec{k}) = \langle \psi_{\vec{k}\alpha} | \hat{\vec{p}} | \psi_{\vec{k}\alpha^\prime}\rangle $ are the matrix elements of the momentum operator $\hat{\vec{p}}$, known as the velocity matrix elements. The pump pulse is described by the vector potential $\vec{A}_\mathrm{p}(t)$ within the dipole approximation.

To calculate pump-probe ARPES intensities, we combine the light-matter coupling entering the Hamiltonian~\eqref{eq:tdham} and the photoemission matrix elements~\eqref{eq:mel_ansatz} with the nonequilibrium steady-state (NESS) approach. For a sufficiently long pump pulse, the trARPES signal for the probe pulse at the center of the pump can be described by the Floquet bands calculated from a purely time-periodic Hamiltonian. Due to electron-electron and electron-phonon scattering, the excited states reach a quasi-thermal state, thus determining the transient occupation of the Floquet bands~\cite{murakami_nonequilibrium_2017}. In this regime, the NESS approach yields a realistic model for the trARPES signal~\cite{schuler_how_2020}.
The steady state reached by the dynamical balance of absorption and a generic type of dissipation is modeled by assuming that the monolayer WSe$_2$ is coupled to a wide-band thermalizing bath, while the excitation is induced by a time-periodic vector potential $\vec{A}_\mathrm{p}(t) = \vec{A}_0 \sin(\omega_\mathrm{p} t)$, $\omega_\mathrm{p} = 2\pi/T_\mathrm{p}$. From these ingredients, we compute the lesser Green's function $G^<_{\alpha \alpha^\prime}(\vec{k},t,t^\prime)$ within the NESS formalism (see methods section). Assuming a sufficiently long probe pulse, the trARPES signal is then obtained~\cite{freericks_theoretical_2009,schuler_theory_2021-1} from
\begin{align}
	\label{eq:trARPES}
	I(\vec{k},E) \propto \mathrm{Im} \sum_{\alpha \alpha^\prime} M_{\alpha}(\vec{k},E)G^<_{\alpha \alpha^\prime}(\vec{k},\omega_\mathrm{pr} - E)M^*_{\alpha^\prime}(\vec{k},E) \ ,
\end{align}
where
\begin{align}
	\label{eq:gles}
	G^<_{\alpha \alpha^\prime}(\vec{k},\omega) &= \frac{1}{T_\mathrm{p}} \int^{T_\mathrm{p}}_0 d t_\mathrm{av}\int^{\infty}_{-\infty} dt_\mathrm{rel}\, e^{i \omega t_\mathrm{rel}} \nonumber \\ &\quad \times G^<_{\alpha \alpha^\prime}\left(t_\mathrm{av} + \frac{t_\mathrm{rel}}{2}, t_\mathrm{av} - \frac{t_\mathrm{rel}}{2}\right) 
\end{align}
and $\omega_\mathrm{pr}$ denotes the frequency of the probe pulse.
The lesser Green's function $G^<_{\alpha \alpha^\prime}(\vec{k},\omega)$ contains both the information on the spectrum and the occupation. If we are only interested in the spectrum, we can compute the retarded Green's function $G^\mathrm{R}_{\alpha \alpha^\prime}(\vec{k},t,t^\prime)$, which is transformed to frequency space by a direct analog of Eq.~\eqref{eq:gles}. 

The only parameters of the NESS approach originate from the bath, which is characterized by a coupling strength $\Gamma$ (which also determines the energy resolution) and an effective temperature $T_\mathrm{eff}$. The occupation of the excited states is controlled by $T_\mathrm{eff}$; we choose typical values $T_\mathrm{eff}\sim 10^4 K$, which yield realistic pump-probe photoemission spectra. 

\subsection*{Orbital character and circular dichroism}

\begin{figure}[t]
\centering\includegraphics[width=\columnwidth]{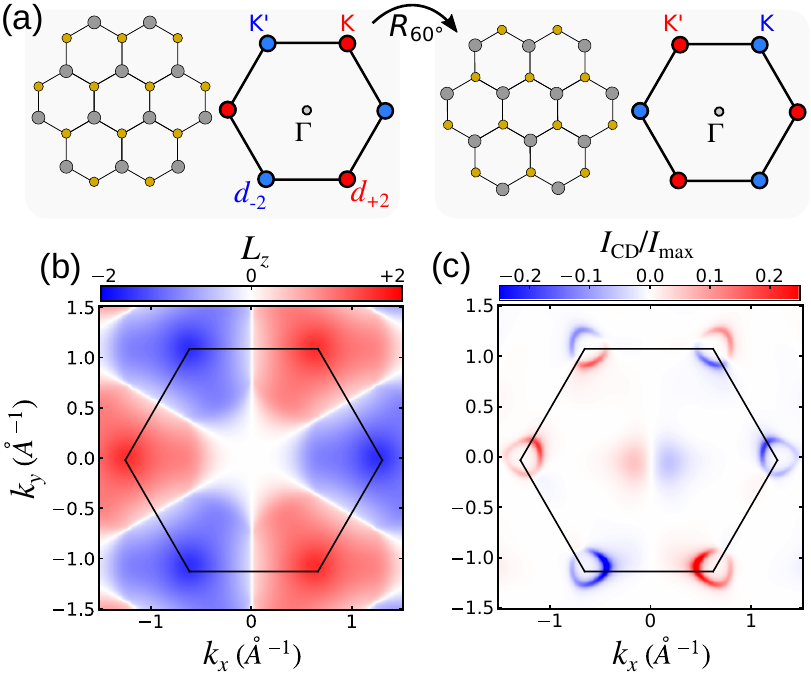}
\caption{\textbf{Orbital angular momentum and circular dichroism.} (a) Top view of the WSe$_2$ monolayer and sketch of the first Brillouin zone. Upon rotating the crystal by 60$^\circ$, the dominant orbital character at K (K$^\prime$) reverse from the magnetic state $d_{+2}$ ($d_{-2}$) to $d_{-2}$ ($d_{+2}$), thus emulating the time-reversal operation.
(b) Out-of-plane component of the orbital angular momentum $L_z$ of the top valence band. 
(c) Calculated CDAD $I_\mathrm{CD}(\vec{k},E) = I_\mathrm{LCP}(\vec{k},E) - I_\mathrm{RCP}(\vec{k},E) $ at $E - E_\mathrm{VBM} = -0.215$~eV for the orientation on the left-hand side of (a). The CDAD for the rotated crystal orientation is qualitatively similar.}
\label{fig:cdad_rot}
\end{figure}

Before presenting the pump-probe spectra, it is instructive to discuss the orbital character of the relevant bands in equilibrium. The Bloch state of top valence band close to the VBM is dominated by the W $d_{z^2}$, $d_{xy}$ and $d_{x^2-y^2}$ orbitals, which form magnetic states at the K, K$^\prime$ points. Therefore, it is convenient to work in the magnetic basis $j \rightarrow \ell, m$, where $\ell, m$ correspond to complex spherical harmonics. In this basis, the Bloch state in the vicinity of the VBM is given by $|\psi^{\mathrm{K,K}^\prime}_{\vec{k}\alpha}\rangle \approx \left[ C_{2,0}(\vec{k}) |\varphi_{\vec{k},2,0}\rangle + C_{2,\pm 2}(\vec{k})  |\varphi_{\vec{k},2,\pm 2}\rangle \rangle  \right] \otimes |\uparrow, \downarrow \rangle$, where $|\varphi_{\vec{k},2,0}\rangle$ ($|\varphi_{\vec{k},2,\pm 2}\rangle$) denotes a Bloch state with $d_{z^2}$ ($d_{\pm 2} = (d_{x^2-y^2} \pm d_{xy})/\sqrt{2}$) orbitals localized at the W sites only. The orbital angular momentum (OAM) of a given Bloch state is reversed by rotating the crystal by 60$^\circ$, which swaps K$\leftrightarrow \mathrm{K}^\prime$ (see illustration in Fig.~\ref{fig:cdad_rot})(a)). Hence, the rotation operation $R_{60^\circ}$ is equivalent to a time-reversal operation~\cite{beaulieu_revealing_2020-1}. 

The OAM of the Bloch state can directly be accessed by circularly polarized probes. In particular, the OAM is reflected in circular dichroism in photoelectron angular distributions ~\cite{razzoli_selective_2017,cho_experimental_2018,schuler_local_2020-1,cho_studying_2021,unzelmann_momentum-space_2021}, which can be understood intuitively by considering both the helicity of light and the self-rotation of the Bloch states~\cite{xiao_berry_2010}. In Fig.~\ref{fig:cdad_rot}(b) we show the $z$ component of the OAM of the top valence band, which qualitatively aligns with the texture of the $d_{\pm 2}$ orbitals. How exactly the OAM and the circular dichroism are related quantitatively depends on details of the experimental geometry and the photon energy. Here we consider the geometry sketched in Fig.~\ref{fig:sketch}(b). In this setup, left/right-hand circular polarized (LCP/RCP) light is defined by the polarization vector $\vec{e}_\mathrm{LCP/RCP} = (\vec{e}_s(\theta) \mp i \vec{e}_p(\theta))/\sqrt{2}$, where $\vec{e}_s(\theta)$ ($\vec{e}_p(\theta)$) denote the direction of $s$ ($p$) polarization with the incidence angle $\theta$. As in our previous experiments~\cite{beaulieu_revealing_2020,beaulieu_unveiling_2021,schuler_polarization-modulated_2021}, we fix $\theta=65^\circ$ and $\hbar \omega_\mathrm{pr}=21.7$~eV. 

In Fig.~\ref{fig:cdad_rot}(v) we present the circular dichroism $I_\mathrm{CD}(\vec{k},E) = I_\mathrm{LCP}(\vec{k},E) - I_\mathrm{RCP}(\vec{k},E)$, calculated from Eq.~\eqref{eq:trARPES} (in absence of a pump pulse), where the polarization vector $\vec{e}_\mathrm{LCP/RCP}$ were inserted in the definition of the matrix elements~\eqref{eq:mel_ansatz}. The circular dichroism in the angular distribution (CDAD) for binding energies close to the VBM is predominantly positive (negative) at K (K$^\prime$), displaying its direct connection to the OAM. The structure of the CDAD is more complex around the top edge of the Brillouin zone, which is due to the specifics of the geometry, as $\theta\ne 0$ breaks the reflection symmetry of the ARPES signal. Since for a 2D system $L_z$ is the only relevant component, we would expect the one-to-one correspondence of the OAM only for normal incidence~\cite{schuler_local_2020-1}. However, even for strongly tilted incidence angle ($\theta=65^\circ$) the CDAD provides an excellent qualitative map of the OAM texture. Therefore, the CDAD in the present geometry is expected to be an excellent marker for the pump-induced modifications of the orbital texture.

\subsection*{Floquet spectra}

Before presenting the CDAD in the presence of the driving pump pulse, we discuss the photodressed electronic structure and associated orbital texture. To this end we have calculated the retarded Green's function $G^\mathrm{R}_{\alpha \alpha^\prime}(\vec{k},\omega)$, including the interaction with the pump pulse via the Hamiltonian~\eqref{eq:tdham}, and the Floquet spectral function
\begin{align}
	\label{eq:apsect}
	A(\vec{k},\omega) = -\frac{1}{\pi}\mathrm{Im}\sum_{\alpha} G^\mathrm{R}_{\alpha \alpha}(\vec{k},\omega) \ ,
\end{align}
which can be seen as a steady-state extension of the density of states. Eq.~\eqref{eq:apsect} is similar to the trARPES intensity~\eqref{eq:trARPES} upon replacing the matrix elements $M_{\alpha}(\vec{k},E)M^*_{\alpha^\prime}(\vec{k},E)\rightarrow \delta_{\alpha \alpha^\prime}$.

\begin{figure}[t]
\centering\includegraphics[width=\columnwidth]{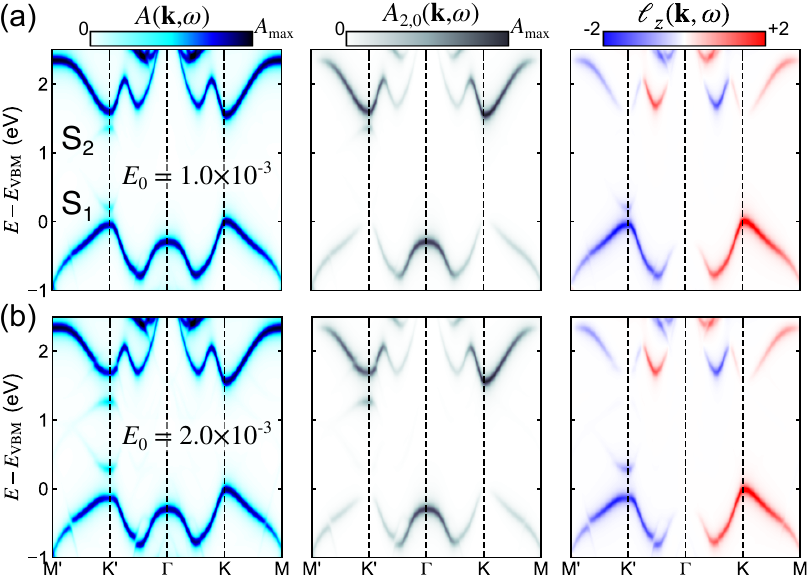}
\caption{\textbf{Floquet spectrum and orbital texture.} Spectral function $A(\vec{k},\omega)$ in the presence of a  driving laser, associated spectral function $A_{2,0}(\vec{k},\omega)$ projected onto $(\ell,m)=(2,0)\equiv d_{z^2}$ orbitals, and OAM texture $\ell_z(\vec{k},\omega)$ defined by Eq.~\eqref{eq:oam_text}, for (a) $E_0=1.0\times 10^{-3}$~a.u., and  (a) $E_0=2.0\times 10^{-3}$~a.u. peak pump field strength. The pump photon energy is $\hbar \omega_\mathrm{p}=1.4$~eV and the pump field is left-handed circularly polarized.} 
\label{fig:spectrum}
\end{figure}

We consider sub-gap (band gap in our model $E_\mathrm{gap}=1.52$~eV) pumping $\hbar \omega_\mathrm{p}=1.4$~eV for two reasons. (i) Red-detuned pumping is the most direct path to realizing the topologically nontrivial state in the conduction band manifold as discussed in ref.~\cite{claassen_all-optical_2016-1}. (ii) Absorption is strongly suppressed, which is crucial for avoiding laser-heating effects. While there would be pronounced resonant exciton features for the monolayer, the strong screening in the bulk sample reduces the exciton binding energy and oscillator strength. Hence, in this case, we neglect excitonic effects. The peak field strength of the pump pulse is chosen in the range $E_0 = 1\times 10^{-3}$ a.u. to $E_0 = 2\times 10^{-3}$ a.u., which corresponds to peak intensity $I_0 \sim 3.5 \times 10^{10}$~W/cm$^2$ to $I_0 \sim 1.4 \times 10^{11}$~W/cm$^2$. This field strength range is well within the experimentally feasible regime.

Fig.~\ref{fig:spectrum} shows the Floquet spectral function~\eqref{eq:apsect} for LCP pumps at two different field strength. For clarity of the discussion, we ignore spin-orbit coupling (SOC) at this point.
The driven band structure exhibits only subtle changes in this regime, apart from small gap openings in the first conduction band and top valence band at K$^\prime$. Note the optical absorption selection rules suppress photodressing at K points.  The most striking features are the side band features S$_1$ (S$_2$) directly above (below) the valence (conduction) band. These side bands separate further from the main bands upon increasing the pump strength, indicating light-induced orbital hybridization. To elucidate the orbital texture of these Floquet features, we have projected the Floquet spectral function~\eqref{eq:apsect} onto the Wannier functions $|\varphi_{\vec{k},\ell, m}\rangle$ in angular momentum basis. The projected spectral function $A_{\ell,m}(\vec{k},\omega)$ thus describes the energy- and momentum resolved orbital weight of $|\varphi_{\vec{k},\ell, m}\rangle$. The Floquet spectrum for $(\ell,m)=(2,0)\equiv d_{z^2}$ is presented in the middle panels in Fig.~\ref{fig:spectrum}. The weight of the $d_{z^2}$ orbital is dominant for the first conduction band at K/K$^\prime$ and for the side band S$_2$. 

\begin{figure}[t]
\centering\includegraphics[width=\columnwidth]{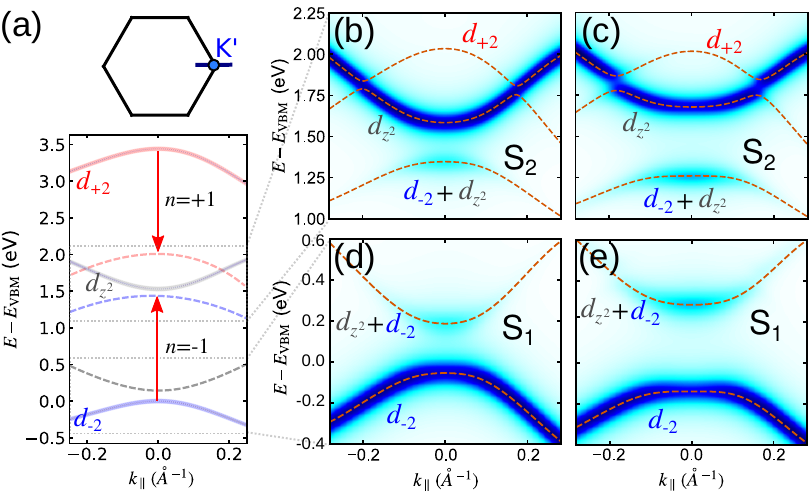}
\caption{\textbf{Orbital character of Floquet side bands.} (a) Sketch of the band structure and dominant orbitals near K$^\prime$ along the sketched path in reciprocal space. The dashed lines are periodic replicas of the bottom conduction band (grey), top valence band (blue), and a higher conduction band (red), respectively. 
(b) Zoom into the region close to the side band features S$_2$ in Fig.~\ref{fig:spectrum} for $E_0=1 \times 10^{-3}$~a.u., and (c) $E_0=2\times 10^{-3}$~a.u. peak field strength. (d) Zoom into the region of $S_1$ for $E_0=1.0\times 10^{-3}$~a.u., and (d) $E_0=2 \times 10^{-3}$~a.u. pump amplitude.
The dashed lines represent the band structure of the downfolded model.} 
\label{fig:inset}
\end{figure}

The orbital-projected spectral function $A_{\ell,m}(\vec{k},\omega)$ also allows to define the spectral density of OAM via
\begin{align}
	\label{eq:oam_text}
	\ell_z(\vec{k},\omega) = \sum_{\ell,m} m A_{\ell,m}(\vec{k},\omega) \ .
\end{align}
Note that Eq.~\eqref{eq:oam_text} is an approximation ignoring non-local contributions of self-rotating Bloch states~\cite{caruso_chirality_2021}. Nevertheless, including the local orbital contributions only has been shown to be accurate for WSe$_2$ in refs.~\cite{cho_experimental_2018,cho_studying_2021}. The OAM texture~\eqref{eq:oam_text} is shown in the right panels in Fig.~\ref{fig:spectrum}. At K/K$^\prime$ the OAM reflects the magnetic orbital character as sketched in Fig.~\ref{fig:cdad_rot}(a). Interestingly, the side band S$_1$ inherits the OAM texture of the top valence band. 

The character of the involved orbitals and their hybridization can be further pinned down by projecting into the relevant subset of orbitals. This is achieved by the downfolding technique (we use the quadratic muffin-tin orbital method from ref.~\cite{zurek_muffin-tin_2005}). We start from Floquet Hamiltonian
\begin{align}
	\label{eq:floqham}
	\mathcal{H}_{\alpha n, \alpha^\prime n^\prime}(\vec{k}) = \frac{1}{T_\mathrm{p}}\int^{T_\mathrm{p}}_0 dt\, e^{i (n-n^\prime) \omega_\mathrm{p} t}H_{\alpha\alpha^\prime}(\vec{k},t) - n \omega_\mathrm{p}\delta_{\alpha \alpha^\prime}\delta_{n n^\prime} \ ,
\end{align}
where $n, n^\prime$ label the photon ladder. Diagonalizing the Floquet Hamiltonian~\eqref{eq:floqham} yields the photodressed bands; it is also used to compute the NESS Green's functions~\eqref{eq:gles}. To understand the orbital texture of S$_2$, we downfold onto the subspace $\{(d_{z^2},n=0), (d_{-2}, n=-1), (d_{+2},n=+1)\}$ (see Fig.~\ref{fig:inset}(a) for an illustration). This set corresponds to the original conduction band ($d_{z^2},n=0$), the first side band of the top valence band upon absorbing one photon ($d_{-2},n=-1$), and the first side band of a higher-lying conduction band ($d_{+2}, n=+1$) upon emitting photon. Downfolding the full Floquet Hamiltonian~\eqref{eq:floqham} into this subspace yields an effective $3\times 3$ Hamiltonian; the corresponding structure is shown (as dashed lines) in Fig.~\ref{fig:inset}(b),(c). The excellent match of the first-principle Floquet spectrum and the downfolded bands underline that the orbital texture is captured by these few orbitals. While S$_2$ originates from the $d_{-2}$-dominated valence band, the top of the side band acquires $d_{z^2}$ character. The two other bands passing through the original conduction band exhibit orbital inversion stemming from the conduction band and the side-band of the higher-lying $d_{+2}$ crossing. Consistent with ref.~\cite{claassen_all-optical_2016-1}, these two bands -- viewed as eigenstates of a static Hamiltonian -- form a Chern insulator state with Chern number $C=1$. 

The orbital texture of S$_1$ (Fig.~\ref{fig:inset}(a)) is simpler: the photodressed top valence band retains its $d_{-2}$ character, while the first side band of the conduction band (minus a photon) acquires some $d_{-2}$ character in addition to its predominant $d_{z^2}$ weight (Fig.~\ref{fig:inset}(d),(e)). This transfer of orbital weight is compensated by the gain of $d_{z^2}$ character of S$_2$.

\begin{figure*}[t]
\centering\includegraphics[width=\textwidth]{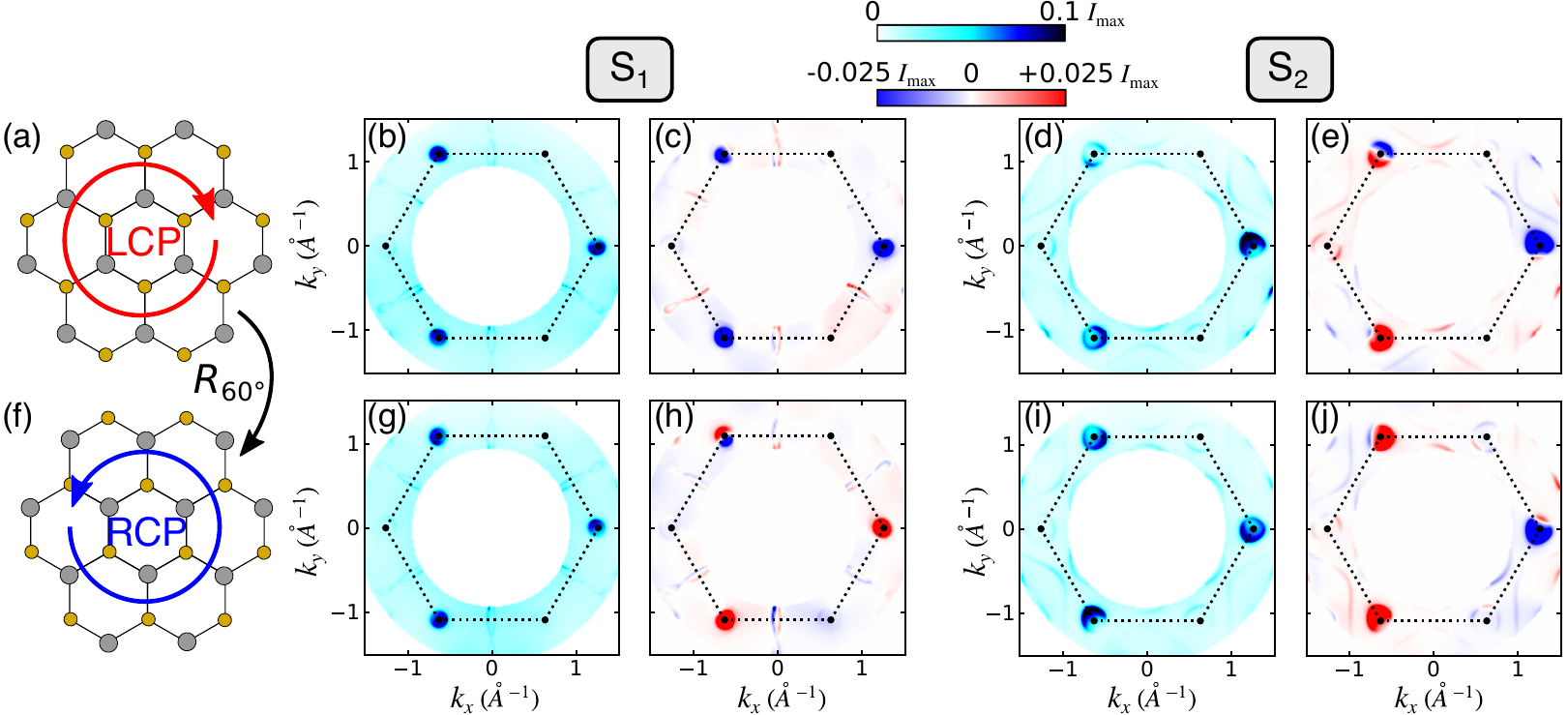}
\caption{\textbf{Intensity and circular dichroism of Floquet side bands.} (a) Sketch of the sample orientation and LCP pump polarization for the spectra in (b)--(e). (b) Photoemission intensity at a fixed-energy cut through S$_1$; (c) corresponding CDAD. (d),(e): analogous plots of intensity and CDAD for a fixed-energy cut through S$_2$. (f) Sketched of rotated sample and swapped pump helicity, which is the setup for the spectra in (g)--(j). The intensity for S$_1$/S$_2$ is shown in (g)/(i), while the CDAD is presented in (h)/(j). The binding energy $E$ is chosen to maximize the intensity; the pump field strength is $E_0=2\times 10^{-3}$ a.u., corresponding to Fig.~\ref{fig:spectrum}(b).} 
\label{fig:floq_cdad}
\end{figure*}

Although the Floquet topological state is realized in the conduction band, the steady-state orbital weight -- which determines the observable photoemission intensity -- follows the original (trivial) $d_{z^2}$ band. Hence, the total occupation-weighted Berry curvature is vanishingly small, since the Berry curvature of the upper and lower band are opposite. Evidencing the light-induced topological state by a quantized Hall response (which probes the global - momentum integrated - topological nature of the system) is thus not feasible. The orbital weight transfer in the side bands S$_1$ and S$_2$, on the other hand, is a fingerprint of the topological state since the specific hybridizations and the resulting orbital textures are tied to the topological state. Therefore, one needs to have an observable which is sensitive to the local (in momentum-space) topological character of the band structure. As explained in the Introduction, this can be achieved by looking at the photoemission intensity modulation upon swapping the helicity of the ionizing probe pulse, \textit{i.e.} by measuring the CDAD.

\subsection*{Pump-probe circular dichroism}

Exploiting the helicity of the circularly polarized probe pulse, the orbital texture of the side bands can be accessed. In particular, we consider the CDAD as a direct probe of magnetic properties and orbital symmetry~\cite{schuler_local_2020-1}. We compute the photoemission spectrum from Eq.~\eqref{eq:trARPES} using the full matrix elements~\eqref{eq:mel_ansatz}. At equilibrium, this approach reproduces the CDAD in Fig.~\ref{fig:cdad_rot}(b), while in the presence of the driving laser pulse, we can directly study the CDAD of the Floquet features and thus access its orbital texture.

We focus on the orbital texture of the side bands S$_{1,2}$. Fig.~\ref{fig:floq_cdad} presents the pump-probe photoemission intensity and the CDAD (measured by swapping the ellicity of circularly polarized probe pulse), for the driven WSe$_2$ sample. Due to optical selection rules, for the sample orientation in Fig.~\ref{fig:floq_cdad}(a), the LCP pump gives rise to Floquet side bands only in the vicinity of the K$^\prime$ valleys, as in Fig.~\ref{fig:spectrum}. Hence, the intensity (Fig.~\ref{fig:floq_cdad}(b), (d)) exhibits a threefold symmetric pattern, with pronounced features at the K$^\prime$ points. Inspecting the CDAD of S$_1$ (Fig.~\ref{fig:floq_cdad}(c)), we observe negative CDAD, while the CDAD of S$_2$ shows, when averaged around the K$^\prime$ valleys, positive (negative) CDAD for $k_x > 0$ ($k_x < 0$). 

Now we consider the sample rotated by 60$^\circ$ and pumped by RCP light (Fig.~\ref{fig:floq_cdad}(f)). Since this sample rotation is equivalent to a time-reversal operation, also the pump polarization needs to be time-reversed for the side bands to appear at the same position (same valleys) in momentum space. While the intensity of S$_1$ is very similar, the valley-averaged CDAD changes its sign. This behavior is a clear indication of the OAM reversing its sign, which is consistent with the acquired $d_{-2}$ ($d_{+2}$) character of the side band of the original conduction band. Indeed, if we assume that only the $d_{\pm 2}$ orbital contributes to the matrix element~\eqref{eq:mel_ansatz}, one finds for the corresponding CDAD $I^{(2,+2)}_\mathrm{CD}(\vec{k}, E) \approx -I^{(2,-2)}_\mathrm{CD}(\vec{k}, E)$ on completely general grounds (see Supplementary Note). This equality holds for normal incidence.

In contrast to the behavior of S$_1$, the CDAD of S$_2$ does not show an overall sign reversal (except for some subtle intra-valley modifications). This is consistent with $d_{z^2}$ orbital character, which is not affected by the time-reversal operation upon rotating by 60$^\circ$. Furthermore, the sign change along $k_x$ is indicative of $d_{z^2}$ character. As shown in the Supplementary Note, the CDAD for a Bloch state comprised of only the $d_{z^2}$ orbital obeys $I^{(2,0)}_\mathrm{CD}(-k_x,k_y,E) = -I^{(2,0)}_\mathrm{CD}(k_x,k_y,E)$.

In summary, the CDAD of the side bands directly reflects the orbital character acquired by the hybridization of the valence band and a replica of the conduction band (S$_1$) or the hybridization of the conduction band a copy of the valence band (S$_2$).

\subsection*{Robustness of dichroic markers against dissipation}

\begin{figure*}[t]
\centering\includegraphics[width=\textwidth]{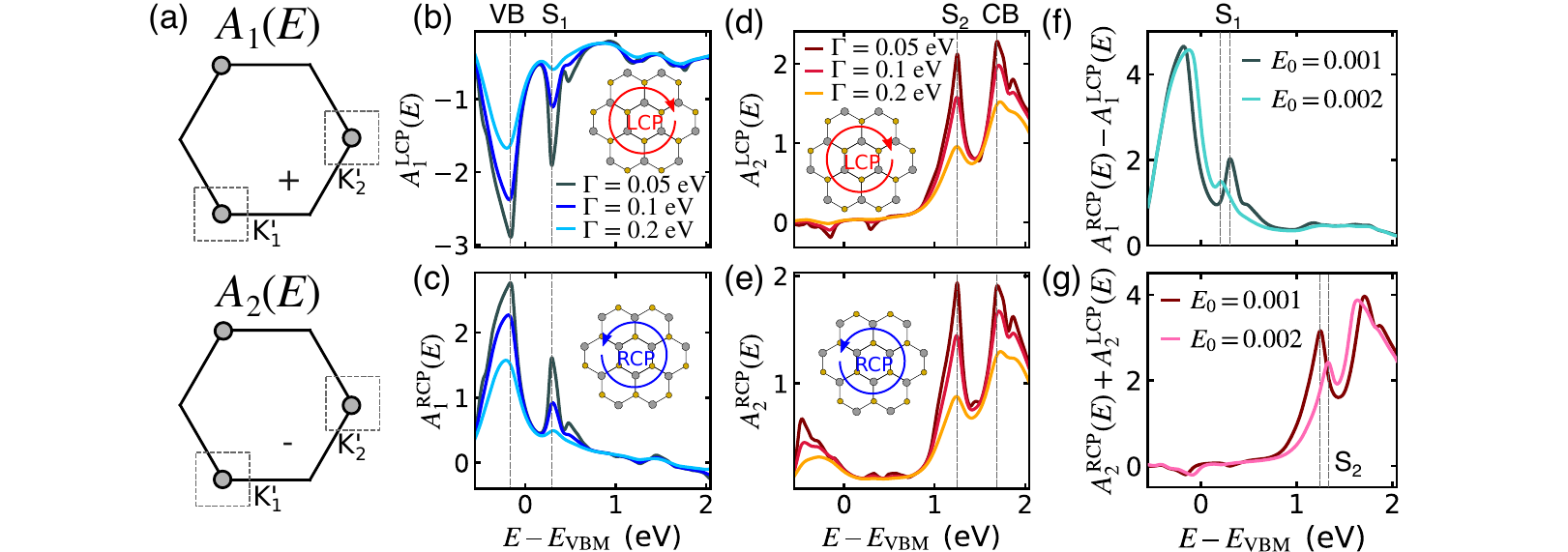}
\caption{\textbf{Dichroic markers and robustness against dissipation.} (a) Illustration of the definition of the dichroic markers $A_{1,2}(E)$ for the side bands S$_{1,2}$. (b) dichroic marker $A_1(E)$ for the orientation and LCP pump polarization as in Fig.~\ref{fig:floq_cdad}(a), and (c) for orientation and RCP pump polarization as in Fig.~\ref{fig:floq_cdad}(f). (d), (e) dichroic marker $A_2(E)$ for crystal orientation as in (b), (c). The pump field strength is $E_0=2\times 10^{-3}$\, a.u. in (b)--(e), and the signal for different bath coupling strength $\Gamma$ is shown. We also indicate the position of the valence band (VB), side band band ($S_{1,2}$), and conduction band (CB) peaks. (f) Time-reversal difference of the marker $A_1(E)$ for $\Gamma = 0.1$~eV for different field strength of the pump. The dashed lines in (b), (c), (f) indicate the position of the side band S$_1$, while they indicate S$_2$ in (d), (e), (g).}
\label{fig:valleycd}
\end{figure*}

The CDAD of the Floquet side bands and their behavior under rotating the crystal (effective time-reversal operation) allows us to define dichroic markers, which are expected to be robust and thus measurable in experiments. In the present geometry, the CDAD for $k_y \le 0$ is less affected by extrinsic (geometric) matrix-element effects within the K$^\prime$ valleys; hence, we focus on the CDAD around K$^\prime_{1,2}$ as illustrated in Fig.~\ref{fig:valleycd}(a). We define the valley-integrated intensity $I_\mathrm{av}(\mathrm{K}^\prime_{1,2},E)$ and CDAD $I_\mathrm{CD}(\mathrm{K}^\prime_{1,2},E)$, where the momentum integration is performed over the dashed squares in Fig.~\ref{fig:valleycd}(a). Because the same (opposite) sign of the CDAD around K$^\prime_{1,2}$ is expected for S$_{1}$ (S$_2$), we introduce the dichroic markers
\begin{align}
	\label{eq:defa1}
	A_1(E) &= \frac{I_\mathrm{CD}(\mathrm{K}^\prime_{1},E)}{I_\mathrm{av}(\mathrm{K}^\prime_{1},E)} + \frac{I_\mathrm{CD}(\mathrm{K}^\prime_{2},E)}{I_\mathrm{av}(\mathrm{K}^\prime_{2},E)} \ ,\\
	\label{eq:defa2}
	A_2(E) &= \frac{I_\mathrm{CD}(\mathrm{K}^\prime_{1},E)}{I_\mathrm{av}(\mathrm{K}^\prime_{1},E)} - \frac{I_\mathrm{CD}(\mathrm{K}^\prime_{2},E)}{I_\mathrm{av}(\mathrm{K}^\prime_{2},E)} \ .
\end{align}
Given the expected momentum dependence of the dichroism for $d_{z^2}$ and $d_{-2}$/$d_{+2}$ orbitals explained above, a peak in $A_1(E)$ ($A_2(E)$) would reflect a state with strong $d_{-2}$/$d_{+2}$ ($d_{z^2}$) character. These markers hence allow reducing complex momentum-resolved CDAD into integrated quantities that are directly reflecting the orbital character of a given equilibrium or light-induced state. The normalization to the intensity allows for a direct comparison under different pump and dissipation conditions. In Fig.~\ref{fig:valleycd}(c),(d) we present the marker~\eqref{eq:defa1} for both crystal orientations. The dichroic marker shows two distinct peaks that correspond to the states with pronounced OAM. The first peak originates from the top valence band with its almost unchanged $d_{-2}$ character, while the second peak reflects the acquired $d_{-2}$ character of S$_1$ (see Fig.~\ref{fig:spectrum}). This peak is very pronounced for low to moderate dissipation and remains visible even for a bath coupling strength of $\Gamma = 0.2$~eV. The corresponding energy smearing ($\sim 80$\,meV) is larger than the spectral resolution that can be achieved in trARPES setups~\cite{}. Note that the sign of peaks changes upon applying the time-reversal operation, which reverses the OAM of the magnetic orbitals $d_{-2}\leftrightarrow d_{+2}$. 

Inspecting the marker~\eqref{eq:defa2} sensitive to sign changes along $k_x$, we see that most of the features near the VBM cancel, while the region close to the conduction band is highlighted. There is a pronounced peak corresponding to S$_2$ and to the lowest conduction band. Both peaks are attributed to $d_{z^2}$ orbital character. The S$_2$ peak is even more robust against dissipation than the S$_1$ peak. 

The sign reversal of the marker~\eqref{eq:defa1} and absence thereof in the marker~\eqref{eq:defa2} upon crystal rotation prompts us to symmetrize or anti-symmetrize, respectively, the energy-resolved dichroic markers. To highlight the reversal of the OAM, we calculated the difference of the dichroic marker~\eqref{eq:defa1} upon effective time-reversal operation, shown in Fig.~\ref{fig:valleycd}(f). Similarly, the calculated average upon effective time-reversal operation highlights the nonmagnetic orbitals and $d_{z^2}$ in particular (Fig.~\ref{fig:valleycd}(g)). As Fig.~\ref{fig:valleycd}(f), (g) demonstrate, for moderate strength of the dissipation $\Gamma=0.1$~eV the S$_{1,2}$ peaks remain visible even an amplitude of the pump pulse as weak as $E_0=10^{-3}$ a.u.. 

Note that we have excluded spin-orbit coupling (SOC) in the results presented here for clarity of the discussion. Neglecting the SOC is justified by the pronounced spin splitting of the top valence band. In the sub-gap pumping regime, the split-off Floquet side band is clearly separated from the conduction band and thus far off-resonant. The hybridization of the Floquet side bands and the valence and conduction band is suppressed, and the photoemission intensity of the side bands is about two orders of magnitude lower. Hence, our simulations without SOC are predictive. This is underpinned by additional calculations including SOC in the Supplementary Note.

\section*{Discussion}

In summary, we presented how the local topological properties and orbital texture of pumped $2H$-WSe$_2$ manifest in circular dichroism in time-resolved ARPES. Combining first-principle light-matter coupling and a highly-accurate model for the photoemission matrix elements allowed us to predict the distinct dichroic features under realistic experimental conditions. In this setup, the manifestation of the topological state is masked by the quasi-thermal population of the Floquet bands. However, the orbital texture of the side bands -- which is intimately connected to the induced topology -- gives rise to distinct CDAD. Defining robust observables that are sensitive to the orbital character, we showed that we can access the light-induced orbital texture, even for strong dissipation and relatively weak pump fields. 

Focusing on the CDAD as a hallmark of the Floquet state also, in principle, removes LAPE contributions. In the present geometry, LAPE gives rise to photoemission intensity overlapping (and also interfering~\cite{park_interference_2014}) with Floquet side bands. This LAPE enhancement of the photoemission intensity in this spectral region cancels out in the dichroic observables, since it depends only weakly on the helicity of the probe pulse in the regime of XUV probe pulses. 
Therefore -- supported by our predictive calculations -- measuring the pump-probe CDAD is an ideal platform to evidence the emergence of light-induced topological state in $2H$-WSe$_2$ in forthcoming experiments. Indeed, while time-resolved CD-ARPES in the XUV spectral range (allowing extending the photoemission horizon up to Brillouin zone edges (e.g. K/K')) has never been archived yet, all the technological building blocks allowing to performed these experiments have been demonstrated independently and are currently being put together by some experimental groups. The last step towards time-resolved CD-ARPES in the XUV spectral range is to achieved polarization control of high-repetition-rate XUV source, which is not fundamentally different to what has been already demonstrated for femtosecond XUV source pumped by Ti:Sa drivers, operating at moderate repetition-rate~\cite{vodungbo_polarization_2011,willems_probing_2015}. This is allowing us to believe that the methodology described in this manuscript is experimentally within reach.

In addition, recent static CDAD ARPES experiments also showed the signatures of Berry curvature from Weyl semimetals~\cite{unzelmann_momentum-space_2021}; time-resolved CDAD ARPES would thus also be a powerful tool to evidence Floquet-Weyl states~\cite{hubener_creating_2017} and other light-induced topological phenomena.

\section*{Acknowledgments}

The calculations have been performed at the Merlin6 cluster at the Paul Scherrer Institute.
M.S. thanks the Swiss National Science Foundation SNF for its support with an Ambizione grant (project No.~193527). 

\section*{Methods}

\subsection*{Wannier functions and photoemission matrix elements}

We computed the electronic structure of monolayer WSe$_2$ using the \textsc{Quantum Espresso} package~\cite{giannozzi_quantum_2009}. We used the PBE functional and norm-conserving pseudopotentials from the \textsc{PseudoDojo} project~\cite{van_setten_pseudodojo_2018}. We constructed symmetry-adapted maximally localized Wannier functions to retain the $d$ and $p$ orbital character associated to the W and Se sites, respectively. More details on the calculations and extracting the velocity matrix elements $\vec{v}_{\alpha \alpha^\prime}(\vec{k})$ can be found in ref.~\cite{schuler_gauge_2021-1}. 

The atomic photoemission matrix elements~\eqref{eq:mel_ansatz} are computed from the Wannier functions $\phi_{j}(\vec{r})$. We approximate the real-space dependence by
\begin{align}
  \label{eq:pwfs}
  \phi_j(\vec{r}) = R_j(r)X_{\ell_j m_j}(\hat{\vec{r}}) \ ,
\end{align}
where $X_{\ell m}(\hat{\vec{r}})$ denotes the real spherical harmonics. We assume that all $d$ ($p$) orbitals possess the same radial dependence: $R_j(r) = R_d(r)$ ($R_j(r) = R_p(r)$).

Following ref.~\cite{schuler_polarization-modulated_2021}, we compute the photoemission matrix elements by (i) switching to length gauge by replacing $\hat{\vec{p}} = -i [\hat{H},\hat{\vec{r}}]$, where $\hat{\vec{r}}$ is the dipole operator, and (ii) approximating the final states by plane waves. We obtain
\begin{align}
	\label{eq:mel_app}
	M_\alpha(\vec{k},E) &= -i (E - \varepsilon_\alpha(\vec{k})) \sum_j C_{j\alpha}(\vec{k}) e^{i \gamma_j} e^{-i \vec{p}\cdot \vec{t}_j} \nonumber \\ &\quad\times \int d \vec{r}\, e^{-i\vec{p}\cdot \vec{r}} \vec{e} \cdot \vec{r} R_j(r)X_{\ell_j m_j}(\hat{\vec{r}}) \ ,
\end{align} 
where $\vec{p}$ is the three-dimensional photoelectron momentum vector defined by $\vec{p}_\parallel = \vec{k}$ and $\vec{p}^2/2 = E + V_0$ ($V_0$ is the inner potential), and where $\vec{t}_j$ denote the positions of the W and Se atoms in the unit cell. The probe polarization is denoted by the complex unit vector $\vec{e}$. The integral in Eq.~\eqref{eq:mel_app} is evaluated by expanding the plane waves in real spherical harmonics. This leads to two radial integrals 
\begin{align}
	I^{(\pm)}_j(E) = \int^\infty_0 d r\, r^3 j_{\ell_j \pm 1}(\sqrt{2 E} r)R_j(r) \ ,
\end{align}
where $j_\ell(x)$ denotes the spherical Bessel functions. The energy dependence of the radial integrals $I^{(1,2)}_j(E)$ is weak in the XUV regime, therefore we approximate $I^{(\pm)}_j(E) \approx I^{(\pm)}_j$, which correspond to four ($j=d,p$) independent parameters. Together with the inner potential $V_0$ and the 11 phase factors $\gamma_j$, there are 17 free parameters. We determined these parameters such that the resulting photoemission intensity $I(\vec{k},E)$ for $s$ and $p$ polarized at binding energy $E-E_\mathrm{VBM} = -0.2$~eV fits the experimental signals (see Supplementary Note for a direct comparison).

\subsection*{Floquet NESS formalism}

We assume that each lattice site (independent of the orbital) is coupled to a thermalizing bath within the wide-band limit approximation (WBLA). The effects of the bath on the system is captured by the self-energy, which is then used to calculate the Green's function. It is convenient to express these quantities in the basis of the Floquet Hamiltonian~\eqref{eq:floqham}. In this basis, the retarded part of the self-energy reads
\begin{align}
	\label{eq:sigma_r}
	\Sigma^\mathrm{R}_{\alpha n, \alpha^\prime n^\prime}(\omega) = -i \frac{\Gamma}{2} \delta_{\alpha \alpha^\prime} \delta_{n n^\prime} \ ,
\end{align} 
while the lesser part (which determines the occupation of the Floquet bands) is given by
\begin{align}
	\label{eq:sigma_l}
	\Sigma^<_{\alpha n, \alpha^\prime n^\prime}(\vec{k},\omega) = i \Gamma  \delta_{\alpha \alpha^\prime} \delta_{n n^\prime}
	n_F(\omega-\mu + n \omega_\mathrm{p}) \ .
\end{align}
Here, $n_F(\omega)$ denotes the Fermi function, which also includes the effective temperature $T_\mathrm{eff}$. Representing the self-energies~\eqref{eq:sigma_r}--\eqref{eq:sigma_l} by compact matrices in band-Floquet space, the Floquet Green's function is then determined solving the Dyson equation
\begin{align}
	\label{eq:dyson}
	[\hat{\mathcal{G}}^\mathrm{R}(\vec{k},\omega)]^{-1} = \omega - \hat{\mathcal{H}}(\vec{k}) - \hat{\Sigma}^\mathrm{R}(\omega) 
\end{align}{}
and the Keldysh equation
\begin{align}
	\label{eq:keldysh}
	\hat{\mathcal{G}}^<(\vec{k},\omega) = \hat{\mathcal{G}}^\mathrm{R}(\vec{k},\omega)\hat{\Sigma}^<(\omega) [\hat{\mathcal{G}}^\mathrm{R}(\vec{k},\omega)]^\dagger \ .
\end{align}
Once the Floquet Green's functions have been computed from Eqs.~\eqref{eq:keldysh}--\eqref{eq:keldysh}, the physical Green's function $G^{\mathrm{R},<}(\vec{k};t,t^\prime)$ and the trARPES signal can be computed. Assuming a long probe pulse, we can employ Eq.~\eqref{eq:gles}, which simplifies to
\begin{align}
	G^{\mathrm{R},<}_{\alpha \alpha^\prime}(\vec{k},\omega) = \sum_n \mathcal{G}^{\mathrm{R},<}_{\alpha n, \alpha^\prime n}(\vec{k},\omega -n \omega_\mathrm{p})\chi_n(\omega) \ ,
\end{align}
where $\chi_n(\omega) = 1$ if $-\omega_\mathrm{p}/2 < \omega - n \omega_\mathrm{p} < \omega_\mathrm{p}/2$ and zero otherwise.


%

\end{document}